\documentclass[prl,twocolumn,aps,tightenlines,showpacs]{revtex4}

\usepackage[dvips]{graphicx}

\usepackage{dcolumn}     

\renewcommand{\case}{\frac}

\begin{document}
 
{

\title{Evolution of Nuclear Spectra with Nuclear Forces}

\author{R. B. Wiringa\cite{rbw} and Steven C. Pieper\cite{scp}}

\affiliation{Physics Division, Argonne National Laboratory, Argonne, IL 60439}

\date{\today}
 
\begin{abstract}
We first define a series of $N\!N$ interaction models ranging from very simple
to fully realistic.
We then present Green's function Monte Carlo calculations of light nuclei
to show how nuclear spectra evolve as the nuclear forces are made
increasingly sophisticated.
We find that the absence of stable five- and eight-body nuclei depends
crucially on the spin, isospin, and tensor components of the nuclear force.
\end{abstract}
 

\maketitle

}

A key feature of nuclear structure, of great importance to the universe as
we know it, is the absence of stable five- or eight-body nuclei.
This simple fact is crucial to both primordial and stellar nucleosynthesis.
It leads to a universe whose baryonic content is dominated by hydrogen and
$^4$He, with trace amounts of deuterium, $^3$He, and $^7$Li.
It also enables stars like our sun to burn steadily for billions of years,
allowing time for the evolution of life intelligent enough to wonder about
such issues.

In this Letter we demonstrate that the binding energies, excitation
structure, and relative stability of light nuclei, including the opening of
the $A=5$ and 8 mass gaps, are crucially dependent on the complicated
structure of the nuclear force.
We do this by calculating the energy spectra of light nuclei using a
variety of nuclear force models ranging from very simple to fully realistic,
and observing how features of the experimental spectrum evolve with the
sophistication of the force.
We find that the spin-isospin and tensor forces present in long-range
one-pion-exchange (OPE) are vital, which in turn may allow us to make a closer
connection between nuclear structure and the underlying features of
QCD~\cite{OFUBHM01,KM97}.

Modern nucleon-nucleon ($N\!N$) potentials, such as the
Argonne~$v_{18}$~\cite{WSS95}, CD~Bonn~\cite{MSS96}, Reid93, Nijm~I,
and Nijm~II~\cite{SKTS94}, fit over 4300 elastic $N\!N$ scattering data
with a $\chi^2 \approx 1$.
These potentials are very complicated, including spin, isospin, tensor,
spin-orbit, quadratic momentum-dependent, and charge-dependent terms,
with $\sim$40 parameters adjusted to fit the data.
Despite this sophistication, these potentials cannot reproduce the binding
energy of few-body nuclei like $^3$H and $^4$He without the assistance of a
three-nucleon potential~\cite{NKG00}.
Three-nucleon ($N\!N\!N$) potentials, such as the Tucson-Melbourne~\cite{TM79},
Urbana~\cite{CPW83}, and Illinois~\cite{PPWC01} models,
are also fairly complicated, depending on the positions, spins,
and isospins of all three nucleons simultaneously.
A combination of $N\!N$ and $N\!N\!N$ potentials, such as the
Argonne~$v_{18}$ and Illinois~2 (AV18/IL2), evaluated with exact Green's
function Monte Carlo (GFMC) many-body calculations, can describe the spectra
of light nuclei very well~\cite{PPWC01,PVW02}.

The AV18 potential contains a complete electromagnetic (EM) interaction
and a strong interaction part which is a combination of OPE
and remaining shorter-range phenomenology.
The strong interaction part is written as a sum of 18 operator terms:
\begin{equation}
       v^{\pi}_{ij} + v^{R}_{ij} = \sum_{p=1,18} v_{p}(r_{ij}) O^{p}_{ij} \ .
\label{eq:operator}
\end{equation}
The first eight operators,
\begin{equation}
O^{p=1,8}_{ij} = [1, {\bf\sigma}_{i}\cdot{\bf\sigma}_{j}, S_{ij},
{\bf L\cdot S}] \otimes~[1,{\bf\tau}_{i}\cdot{\bf\tau}_{j}] \ ,
\label{eq:8op}
\end{equation}
are the most important for fitting S- and P-wave $N\!N$ data.
The additional terms include six operators that are quadratic in {\bf L},
three charge-dependent (CD) terms, and one charge-symmetry-breaking (CSB) term.
The radial functions $v_{p}(r)$ have parameters adjusted to fit the
elastic $pp$ and $np$ scattering data for $E_{lab} \leq 350$ MeV,
the $nn$ scattering length, and the deuteron energy.

The Illinois $N\!N\!N$ potentials include a complete two-pion-exchange piece,
three-pion rings, and a shorter-range phenomenological term:
\begin{equation}
   V_{ijk} = V^{2\pi}_{ijk} + V^{3\pi,\Delta R}_{ijk} + V^{R}_{ijk} ~.
\end{equation}
The five Illinois models (IL1-IL5) each have $\sim$3 parameters adjusted to
fit the energies of 17 narrow states in $A \leq 8$ nuclei as evaluated in GFMC
calculations~\cite{PPWC01}.
Subsequent calculations of an additional 10 states in $A=9,10$ nuclei
show that, without readjustment, the AV18/IL2 combination is able to reproduce
27 narrow states with an rms deviation of only 600 keV~\cite{PVW02}.

The AV18/IL2 Hamiltonian is the standard of comparison for this Letter.
We also present previously reported results for AV18 alone, and for a 
simplified, but still fairly realistic, potential called
AV8$^\prime$~\cite{PPCPW97}.
The ``8'' designates the number of operator components, which in this
case means those of Eq.(\ref{eq:8op}).
The standard Coulomb interaction between protons, $V_{C1}(pp)$, is
retained, but smaller EM terms are omitted.
The prime denotes that this potential is not a simple truncation of AV18,
but a reprojection, which preserves the isoscalar average of the strong
interaction in all S and P partial waves as well as in the $^3$D$_1$
wave and its coupling to $^3$S$_1$.
Consequently, the deuteron bound state is virtually identical to that of AV18,
except that the omission of the small EM terms alters the
binding energy from the experimental value of 2.22 MeV to 2.24 MeV.
Details of this reprojection are given in Ref.~\cite{PPCPW97}.
Recently, the AV8$^\prime$ (without Coulomb) was used in a benchmark
calculation of $^4$He by seven different methods, including GFMC,
with excellent agreement between the various results~\cite{KNG01}.

Here we define five new potentials, which are progressively
simpler reprojections of AV8$^\prime$, designated AV6$^\prime$, AV4$^\prime$,
AVX$^\prime$, AV2$^\prime$, and AV1$^\prime$~\cite{fortran}.
The reprojections preserve as many features of elastic $N\!N$
scattering and the deuteron as feasible at each level.
GFMC calculations of $A \leq 10$ nuclei for these simpler
models show how different features of the spectra correlate with
specific elements of the forces.

The AV6$^\prime$ is obtained by deleting the spin-orbit terms from
AV8$^\prime$ and adjusting the potential to preserve the deuteron binding.
The spin-orbit terms do not contribute to S-wave $N\!N$ scattering, and
are the smallest contributors to the energy of $^4$He \cite{KNG01}.
They also do not contribute to scattering in the $^1$P$_1$ channel,
but are important in differentiating between the $^3$P$_{0,1,2}$ channels.
Thus this model does not give a very good fit to $N\!N$ scattering data.
To fix the deuteron, we choose to subtract a fraction of the AV8$^\prime$
spin-orbit potential's radial function from the central potential in the
$ST=10$ channel, adjusting the coefficient to get an energy of 2.24 MeV:
\begin{equation}
  v^c_{10}(AV6^\prime) = v^c_{10}(AV8^\prime) - 0.3~v^{ls}_{10}(AV8^\prime) ~.
\end{equation}
This choice preserves the OPE potential, while the deuteron D-state and
quadrupole moment are barely changed.
Spin and isospin terms are projected from the $v^c_{ST}$ as
in Eq.(30) of Ref.~\cite{WSS95}, while tensor terms remain the
same as in AV8$^\prime$.

The AV4$^\prime$ potential eliminates the tensor terms.
The $^1$S$_0$ and $^1$P$_1$ potentials are unaffected, but the coupling
between $^3$S$_1$ and $^3$D$_1$ channels is gone and the $^3$P$_{0,1,2}$
channels deteriorate further.
The central $ST=10$ potential is again adjusted to fix the deuteron binding:
\begin{equation}
  v^c_{10}(AV4^\prime) = v^c_{10}(AV6^\prime) + 0.8735~v^{t}_{10}(AV6^\prime) ~,
\end{equation}
but now there is no D-state and no quadrupole moment.

Although many aspects of $N\!N$ scattering have been sacrificed at the
AV4$^\prime$ level, such a potential still allows us to differentiate
between the four possible $ST$ channels.
Any further reduction in the operator structure sacrifices this feature.
We consider three such simplifications: 1) AVX$^\prime$, where the
operators are 1 and the space exchange operator
$P^x_{ij} = -\case{1}{4}(1+\sigma_i\cdot\sigma_j+\tau_i\cdot\tau_j
+\sigma_i\cdot\sigma_j\tau_i\cdot\tau_j)$;
2) AV2$^\prime$, with operators 1 and $\tau_i\cdot\tau_j$;
and 3) AV1$^\prime$, which is just a pure central force.
AVX$^\prime$ allows one to differentiate between the spin-isospin weighted
averages of S-wave and P-wave forces by setting
$v^c + v^x = \case{1}{2}(v^c_{01}+v^c_{10})$ and
$v^c - v^x = \case{1}{10}(9v^c_{11}+v^c_{00})$.
In this case, the average of the S-waves gives a deuteron that is bound by
only 0.43 MeV, but the intrinsic repulsion in odd partial waves is retained.
AV2$^\prime$ allows one to differentiate between $^1$S and $^3$S potentials,
analogous to the Malfliet-Tjon (MT) I-III interaction \cite{MT69}, with the
combinations
$v^c + v^\tau = v^c_{01}$ and $v^c - 3 v^\tau = v^c_{10}$.
Finally, AV1$^\prime$ is just the average of $^1$S and $^3$S potentials,
analogous to the MT~V interaction.
The AV2$^\prime$ preserves the correct deuteron binding of 2.24 MeV, but
the AV1$^\prime$ again has a deuteron bound by only 0.43 MeV.
While the MT interactions were intended only for use
in s-shell ($A \leq 4$) nuclei, they have been used in larger systems acting
either in all partial waves, or only in even partial waves~\cite{BLO00}.
Here we treat AV1$^\prime$, AV2$^\prime$, and AVX$^\prime$
as operators acting in all partial waves.

\begin{figure*}[ht!]
\centering
\includegraphics[angle=270,width=7.0in]{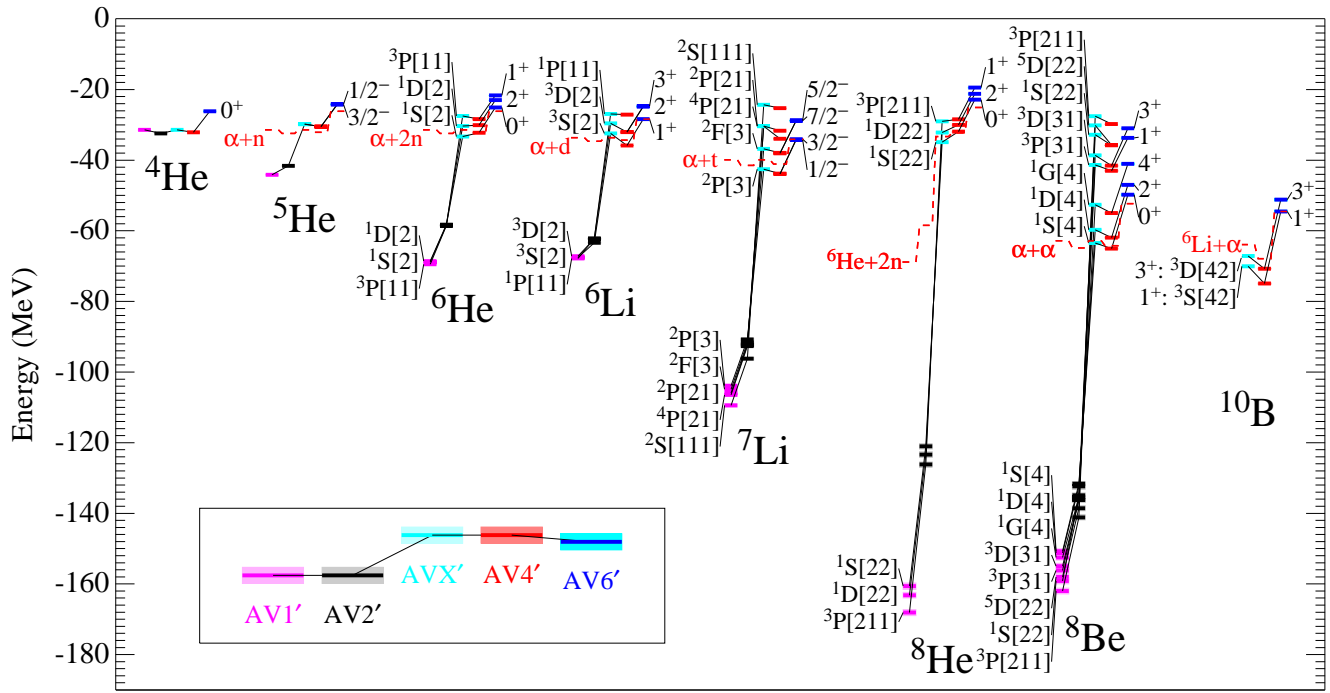}
\caption{Nuclear energy levels for the simpler potential models;
dashed lines show breakup thresholds.}
\label{fig:6down}
\end{figure*}

In the independent-particle (IP) model, nuclear states can be characterized
by quantum numbers $^{2S+1}L[n]$ as well as $J^\pi$, where $L$ and $S$ are
the orbital and spin angular momenta, $[n]$ designates the Young tableau of
the spatial symmetry, and $J^\pi$ are the total angular momentum and
parity~\cite{BM69}.
Realistic interactions down to the AV6$^\prime$ level mix states of different
$^{2S+1}L[n]$, but for the AV4$^\prime$ and simpler interactions there is no
mixing of different values of $S$ and essentially no mixing of $L$ or $[n]$.

\begin{figure*}[ht!]
\centering
\includegraphics[angle=270,width=7.0in]{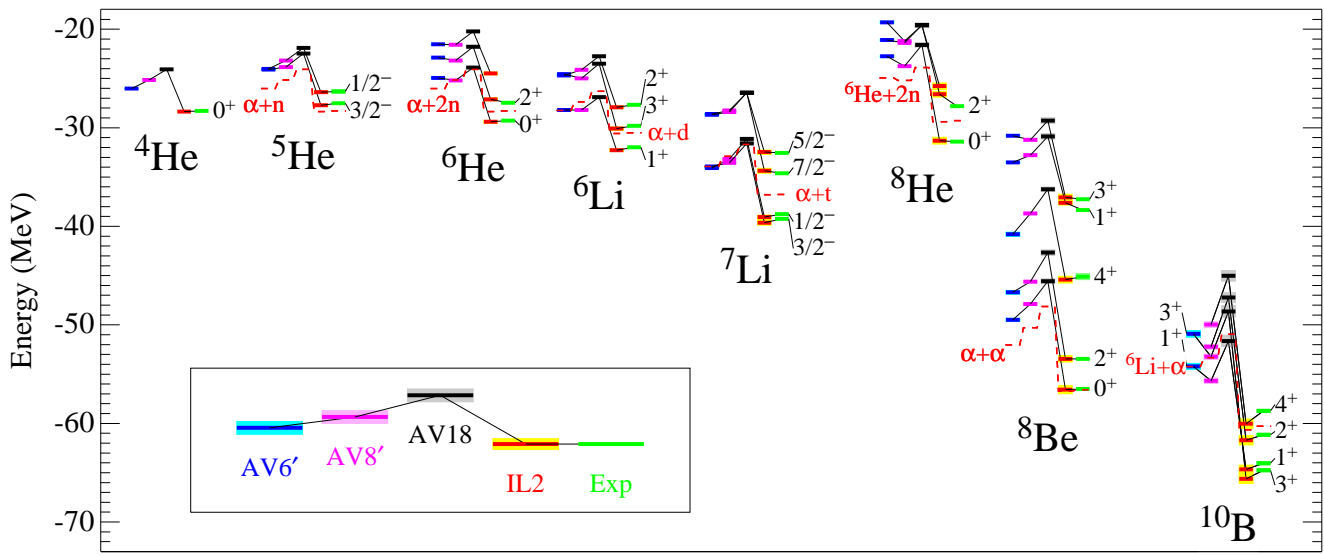}
\caption{Nuclear energy levels for the more realistic potential models;
shading denotes Monte Carlo statistical errors.}
\label{fig:6up}
\end{figure*}

Our many-body calculations are made with the GFMC method,
details of which may be found in Refs.~\cite{PPWC01,PVW02,PPCPW97,WPCP00}.
The GFMC method is in principle exact for the AV8$^\prime$ and simpler
potentials, while for the AV18 and AV18/IL2 models some small parts of
the interaction have to be calculated perturbatively.
We believe the calculation of binding energies for AV18 and AV18/IL2
is accurate to 1--2\%, and better for the simpler models.
Results for 26 $^{2S+1}L[n]$ states in nuclei ranging from $^4$He to
$^{10}$B are shown in Fig.~\ref{fig:6down}.
The Hamiltonians are, from left to right for each nucleus,
AV1$^\prime$, AV2$^\prime$, AVX$^\prime$, AV4$^\prime$, and AV6$^\prime$.
Results for 25 $J^\pi$ states are shown in Fig.~\ref{fig:6up} for
AV6$^\prime$, AV8$^\prime$, AV18, AV18/IL2, and experiment~\cite{results}.

The simple AV1$^\prime$ and AV2$^\prime$ interactions approximately reproduce
the energies of s-shell nuclei ($^3$H is overbound by $\alt 0.6$ MeV).
However, every additional nucleon significantly increases the binding, so
p-shell nuclei are progressively more bound and no mass gaps appear.
With such forces, nuclear matter will not saturate until the repulsive
cores of the interaction start to overlap at many times the empirical density.
For AV1$^\prime$, the strong interaction does not differentiate between
different isospin states, so $^6$He and $^6$Li would have the same energy
except that the Coulomb interaction makes $^6$Li less bound.
Consequently, the $\beta$-stable nuclei in the $4 \leq A \leq 10$ regime
would all be isotopes of helium.
The AV2$^\prime$ avoids this particular problem by preserving separately
the deuteron binding and $^1$S scattering; it also improves
the saturation behavior slightly.
Both models have the curious feature that the spectrum is reversed in order,
i.e., the lowest p-shell eigenstates are the ones that are spatially
{\it least} symmetric.
This lowers the energy by reducing the amount of overlap of the 
repulsive potential cores in the wave function.
Thus the ground state for $^8$Be is a $^3$P[211] state, which has
degenerate spin of $J^\pi=0^+,1^+,2^+$.
The spectra are also very compressed compared to experiment.

The AVX$^\prime$ and AV4$^\prime$ overcome many of the limitations of the
simplest models by preserving the difference between attractive even-
and repulsive odd-partial waves.
Both provide significant saturation, particularly the feature that $A=5$
nuclei are unstable.
However, the $A=8$ mass gap is a more subtle effect, since both these models
predict $^8$Be to have slightly more than twice the binding of $^4$He.
The lowest states are the spatially {\it most} symmetric,
so $^8$Be now has a $^1$S[4] $J^\pi=0^+$ ground state.
The spectrum is also less compressed.
Because the AVX$^\prime$ does not differentiate between $^1$S and $^3$S
interactions, it shares the failing of AV1$^\prime$ in having $^6$He more
bound than $^6$Li, but due to the correct ordering of spatial symmetries,
$^8$Be is much more bound than $^8$He.

The tensor forces in AV6$^\prime$ provide significant
additional saturation compared to the simpler potentials.
This is due to 1) a less attractive $v^c_{10}$ because much of the binding
of the deuteron is now provided through tensor coupling to the $^3$D
channel, and 2) the ability of the tensor interaction with third particles
to change an attractive $^1$S pair into a repulsive $^3$P pair~\cite{FPPWSA96}.
This saturation is sufficiently strong to underbind all the nuclei
with respect to experiment, and it opens the $A=8$ mass gap by
making $^8$Be less than twice as bound as $^4$He.
However, it leaves the $A=6,7$ nuclei with only marginal stability.
The tensor forces begin to mix the $^{2S+1}L[n]$ states so they are no
longer eigenstates, but several sets of states, like the $2^+$ and $3^+$
states in $^6$Li, remain nearly degenerate.

The spin-orbit terms in AV8$^\prime$ provide much more mixing and clearly
break the $J^\pi$ degeneracy, producing a spectrum that is properly ordered
in the $A \leq 8$ nuclei, although the splittings of most spin-orbit
partners are smaller than observed.
The binding energies shift slightly compared to AV6$^\prime$, some up
and some down, with the $A=6,7$ nuclei becoming more stable while
the $A=5,8$ mass gaps are preserved.
Going to the full AV18 interaction decreases the overall binding energies
slightly, because the less attractive higher partial waves in $N\!N$
scattering are now accurately represented, but the relative excitations
are virtually unchanged.
However, the energy differences among
isobaric multiplets, like the $^7$Li,$^7$Be mirror nuclei, are improved
due to the addition of the extra EM, CD, and CSB terms in AV18.

Finally, the addition of the IL2 $N\!N\!N$ potential to AV18 gives a
very accurate description of the light nuclei.
It adds the necessary additional binding that the realistic $N\!N$ potentials
lack and increases the splittings among spin-orbit partners.
It makes $^6$He and $^8$He strong-stable, and
it produces the correct 3$^+$ ground state for $^{10}$B,
where all the simpler models (and other realistic $N\!N$ interactions like
CD Bonn~\cite{navratil}) incorrectly give a 1$^+$ state.

Many of the results from these models can be understood by simply
counting the number of $ST$ pairs, $N_{ST}$, in the IP wave function for 
a given state and applying a weight factor appropriate for the potential.
A good estimate of the relative binding for the AV4$^\prime$ and higher
models is obtained using weights from the OPE potential:
$3N_{10}+3N_{01}-N_{11}-9N_{00}$.
This reflects the approximately equal attraction in $^3$S and $^1$S
potentials, the small repulsion in $^3$P, and the large repulsion in $^1$P.
The values for $^4$He, $^5$He, $^6$He, $^6$Li, $^7$Li, $^8$He, $^8$Be, and
$^{10}$B ground states are 18, 18, 21, 21, 27, 24, 36, and 39, respectively.
With this estimate, the marginal stability of $^5$He and $^8$Be against
breakup and the roughly equal binding of $^6$He and $^6$Li are expected.
It also provides the order of excited states and thus the relative amount of
mixing, e.g., in $^7$Li the 
$^2$P[3], $^4$P[21], and $^2$P[21] states get weights of 27, 21, and 15,
as the number of S-wave pairs decreases going from [3] to [21]
symmetry, and the ratio of $^3$P to $^1$P pairs decreases going from
quartet to doublet spin.

For AV6$^\prime$ and up, the importance of the OPE potential is evident from
its expectation value, which is typically 80\% of $\langle v_{ij} \rangle$
\cite{PPWC01}.
These findings are consistent with the important role of the spin-isospin
interaction in fixing the shell gaps in nuclei~\cite{OFUBHM01},
and support a close connection between nuclear structure and the
underlying features of QCD, particularly the special role of the pion as the
Goldstone boson, and the dominance of spin-isospin and tensor forces in 
$1/N_c$ expansions~\cite{KM97}.

We see from the present studies that purely central nuclear forces are
nonsense for nuclei beyond the s-shell, where it is crucial to incorporate
the difference between attractive even and repulsive odd partial waves.
While a model like AV4$^\prime$ can produce the energy saturation
and clustering that appears in the p-shell, our model calculations
suggest that obtaining the mass gaps at $A=5,8$ {\it and} stable $A=6,7$
nuclei is a very sensitive issue, and may well require {\it both} tensor
and spin-orbit forces as in the AV8$^\prime$ model.
Finally, to get a truly good fit both to the ground state binding energies,
the spin-orbit splittings in the excitation spectra, and (in the case of
$^{10}$B) the ordering of spin states, we need multinucleon forces.

We thank J. L. Friar, D. Kurath, D. J. Millener, and V. R. Pandharipande
for many useful discussions.
The many-body calculations were performed on the parallel computers of the
Mathematics and Computer Science Division, Argonne National Laboratory 
and the National Energy Research Scientific Computing Center.
This work is supported by the U. S. Department of Energy, 
Nuclear Physics Division, under contract No. W-31-109-ENG-38.


\begin{thebibliography}{99}

\bibitem[*]{rbw} Electronic address: wiringa@anl.gov
\bibitem[\dag]{scp} Electronic address: spieper@anl.gov

\bibitem{OFUBHM01}
T. Otsuka {\it et al}.,
Phys. Rev. Lett. {\bf 87}, 082502 (2001).

\bibitem{KM97}
D. B. Kaplan and A. V. Manohar,
Phys. Rev. C {\bf 56}, 76 (1997).

\bibitem{WSS95}
R. B. Wiringa, V. G. J. Stoks, and R. Schiavilla,
Phys. Rev. C {\bf 51}, 38 (1995).

\bibitem{MSS96}
R. Machleidt, F. Sammarruca, Y. Song,
Phys. Rev. C {\bf 53}, R1483 (1996).

\bibitem{SKTS94}
V. G. J. Stoks {\it et al}.,
Phys. Rev. C {\bf 49}, 2950 (1994).

\bibitem{NKG00}
A. Nogga, H. Kamada, and W. Gl\"ockle,
Phys. Rev. Lett. {\bf 85}, 944 (2000).

\bibitem{TM79}
S. A. Coon {\it et al}.,
Nucl. Phys. {\bf A317}, 242 (1979).

\bibitem{CPW83}
J. Carlson, V. R. Pandharipande, and R. B. Wiringa,
Nucl. Phys. {\bf A401}, 59 (1983).

\bibitem{PPWC01}
S. C. Pieper {\it et al}.,
Phys. Rev. C {\bf 64}, 014001 (2001).

\bibitem{PVW02}
S. C. Pieper, K. Varga, and R. B. Wiringa,
nucl-th/0206061, submitted to Phys. Rev. C.

\bibitem{PPCPW97}
B. S. Pudliner {\it et al}.,
Phys. Rev. C {\bf 56}, 1720 (1997).

\bibitem{KNG01}
H. Kamada {\it et al.}, 
Phys. Rev. C {\bf 64}, 044001 (2001).

\bibitem{fortran}
A {\sc fortran} subroutine for AV18, AV8$^\prime$, etc. is available at:
www.phy.anl.gov/theory/fewbody/av18pot.f

\bibitem{MT69}
R. A. Malfliet and J. A. Tjon,
Nucl. Phys. {\bf A127}, 161 (1969).

\bibitem{BLO00}
N. Barnea, W. Leidemann, and G. Orlandini,
Phys. Rev. C {\bf 61}, 054001 (2000).

\bibitem{BM69}
A. Bohr and B. R. Mottelson,
{\it Nuclear Structure Volume I},
(W. A. Benjamin, New York, 1969), Appendix 1C.

\bibitem{WPCP00}
R. B. Wiringa {\it et al}.,
Phys. Rev. C {\bf 62}, 014001 (2000).

\bibitem{results}
A detailed tabulation of the GFMC results is available at:
www.phy.anl.gov/theory/fewbody/avxp\_results.html

\bibitem{FPPWSA96}
J. L. Forest, {\it et al}.,
Phys. Rev. C {\bf 54}, 646 (1996).

\bibitem{navratil}
P. Navr\'atil and W. E. Ormand,
Phys. Rev. Lett. {\bf 88} 152502 (2002);
P. Navr\'atil, private communication.

\end{thebibliography}
\end{document}